\documentclass[pdflatex,sn-mathphys-num]{sn-jnl}

\usepackage{mathrsfs}%
\usepackage{graphicx}%
\usepackage{amsmath,amssymb,amsfonts}%
\usepackage{amsthm}%
\usepackage[title]{appendix}%
\usepackage{xcolor}%
\usepackage{textcomp}%
\usepackage{booktabs}%
\usepackage{bm}%

\newcommand{\dd}{\mathrm{d}}
\newcommand{\ii}{\mathrm{i}}
\newcommand{\eps}{\epsilon}
\newcommand{\Rthree}{{}^{(3)}\mathcal{R}}
\newcommand{\Ricthree}{{}^{(3)}\mathcal{R}^i{}_{j}}
\newcommand{\lap}{\nabla^2}
\newcommand{\diag}{\operatorname{diag}}

\newcommand{\calN}{\mathcal{N}}
\newcommand{\calP}{\mathcal{P}}

\hypersetup{
 pdftitle={Scalar Laplace--Beltrami harmonics and curvature-induced radial mode mixing in Friedmann-like cosmologies},
 pdfauthor={Eduardo Bittencourt, Peter K. S. Dunsby, Bob Osano, and Muzikayise E. Sikhonde},
 pdfkeywords={Laplace--Beltrami operator, cosmological perturbations, Weyl curvature, mode coupling}
}

\begin{document}

\title[Scalar harmonics and curvature-induced radial mode mixing]
{Scalar Laplace--Beltrami harmonics and curvature-induced radial mode mixing in Friedmann-like cosmologies}

\author*[1,2]{\fnm{Eduardo} \sur{Bittencourt}}\email{bittencourt@unifei.edu.br}
\author[2,3,4]{\fnm{Peter K. S.} \sur{Dunsby}}
\author[2,5]{\fnm{Bob} \sur{Osano}}
\author[6]{\fnm{Muzikayise E.} \sur{Sikhonde}}

\affil[1]{\orgdiv{Institute of Physics and Chemistry},
  \orgname{Federal University of Itajub\'a},
  \orgaddress{\street{Av. BPS 1303}, \city{Itajub\'a},
  \postcode{37500-903}, \state{Minas Gerais}, \country{Brazil}}}

\affil[2]{\orgdiv{Department of Mathematics and Applied Mathematics},
  \orgname{University of Cape Town},
  \orgaddress{\street{Rondebosch}, \postcode{7700},
  \city{Cape Town}, \country{South Africa}}}
  
\affil[3]{\orgdiv{Center for Space Research}, \orgname{North-West University}, \orgaddress{\street{Private Bag X6001}, \postcode{2520}, \city{Potchefstroom}, \country{South Africa}}}

\affil[4]{\orgname{South African Astronomical Observatory}, \orgaddress{\street{Observatory}, \postcode{7925}, \city{Cape Town}, \country{South Africa}}}

\affil[5]{\orgdiv{Centre for Higher Education Development},
  \orgname{University of Cape Town},
  \orgaddress{\street{Rondebosch}, \postcode{7701},
  \city{Cape Town}, \country{South Africa}}}

\affil[6]{\orgdiv{Numeracy Centre, Centre for Higher Education Development},
  \orgname{University of Cape Town},
  \orgaddress{\street{Rondebosch}, \postcode{7700},
  \city{Cape Town}, \country{South Africa}}}

\abstract{We develop a harmonic framework for spherically symmetric Friedmann-like geometries with constant scalar curvature but non-zero trace-free Ricci curvature. This preserves Friedmann-type isotropic expansion while introducing a radially varying tidal field and supporting anisotropic stress. We formulate the scalar Laplace–Beltrami operator as a Sturm–Liouville problem and establish its self-adjoint domains and regularity at minimal spheres. Robertson–Walker harmonics are recovered in the maximally symmetric limit. Trace-free curvature terms are expressed as spectral projection kernels. Rotational symmetry preserves angular labels and parity, while the tidal profile couples different radial eigenmodes. A finite-shell calculation displays this off-diagonal structure, verifies consistency between radial representations, and confirms the expected weak-field scaling. A minimally coupled scalar field remains diagonal in the exact basis, indicating operator-dependent mixing rather than mixing solely due to inhomogeneity. We formulate transverse-traceless kernels, whose dynamical application requires a specified model for the supporting anisotropic matter. This converts curvature-induced mode mixing into a spectral problem for perturbations beyond maximally symmetric cosmologies.}

\keywords{Laplace--Beltrami operator, Friedmann-like spacetimes, Sturm--Liouville problem, radial spectral mode mixing}

\maketitle

\section{Introduction}\label{sec:introduction}

Harmonic decomposition is a principal simplifying tool in relativistic cosmology. On maximally symmetric Friedmann--Lema\^itre--Robertson--Walker (FLRW) spatial sections, scalar, vector, and tensor harmonics form invariant sectors of the linearised equations, while amplitudes associated with different Laplace--Beltrami eigenvalues evolve independently \cite{Harrison1967,kodama_sasaki}. At the geometric level, this separation relies not merely on the constancy of the spatial Ricci scalar, but on the stronger Einstein-space condition that the 3-dimensional Ricci tensor is proportional to the induced metric on the constant-time hypersurfaces.

A useful way to relax the FLRW assumptions is to retain a single scale factor and isotropic expansion while allowing the intrinsic spatial curvature to be anisotropic. Perturbation theory on such backgrounds has previously been developed for spatially homogeneous geometries with a preferred direction \cite{zlosnik2011,pereira2015anisotropic,franco2017anisotropic}. The setting considered here is complementary: each spatial section is spherically symmetric but radially inhomogeneous. Rotational symmetry is therefore preserved, whereas radial homogeneity is not.

More specifically, the spatial metric considered below has constant scalar curvature but a nonvanishing trace-free Ricci tensor. In three dimensions, the Weyl tensor vanishes identically, so the Riemann tensor is completely determined by the Ricci tensor. Moreover, an Einstein three-metric locally has constant sectional curvature \cite{besse}. The trace-free Ricci tensor is therefore the geometric obstruction to maximal symmetry in the present setting. When this spatial metric is embedded in the separable four-dimensional geometry, the isotropic part of the Einstein equations retains the Friedmann form, whereas the trace-free spatial curvature gives rise to a nonzero four-dimensional Weyl tensor and requires a supporting background anisotropic stress. This is the precise sense in which the geometry is \emph{Friedmann-like}: its scale factor obeys Friedmann-type equations, but its spatial sections are not those of an FLRW spacetime. Related shear-free, irrotational, geodesic anisotropic-fluid cosmologies were studied in Ref.~\cite{mcmanus_coley1994}, while the particular realisation considered here and its scalar perturbations were developed in Refs.~\cite{bittencourt2014,santos2015}.

The perturbative significance of this departure from maximal symmetry is that the standard FLRW separation into scalar, vector, and tensor sectors is no longer automatic. In an FLRW background, the relevant spatial differential operators can be rearranged without producing direction-dependent curvature terms, allowing the three sectors to evolve independently at linear order. A related failure of the standard FLRW scalar-vector-tensor decoupling occurs in general Lema\^itre--Tolman--Bondi backgrounds \cite{clarkson2009ltb}. In the present geometry, commutators of spatial covariant derivatives instead involve the trace-free Ricci tensor. As a result, curvature-dependent terms are not simultaneously diagonal in the usual harmonic representation and require an explicit spectral projection. This obstruction does not, by itself, imply a dynamical conversion between scalar and tensor degrees of freedom. Spherical symmetry still preserves the angular quantum numbers and parity, so the resulting mode coupling is confined to the radial sector.

The scalar perturbation analysis of Ref.~\cite{santos2015} identified this curvature-induced coupling, but left its spectral structure unresolved. In particular, the exact radial harmonics and the dependence of the coupling coefficients on the radial wave numbers were not determined. The purpose of the present work is to complete that analysis. We construct the scalar Laplace--Beltrami harmonics of the spatial geometry, derive their radial behaviour, and represent the trace-free curvature terms as a kernel acting on the radial spectrum. This formulation makes the radial mode mixing explicit and also separates the consequences of the background geometry from those that depend on the dynamics of the anisotropic matter source. The problem therefore belongs naturally to the spectral theory of Riemannian manifolds and singular Sturm--Liouville operators \cite{chavel,berard,reed_simon,zettl}. It must reproduce the Robertson--Walker bases of Kalnins and Miller when the trace-free Ricci tensor vanishes \cite{kalnins_miller}. For generic nonzero values of geometric parameters, the areal-coordinate radial equation is a five-point Fuchsian equation \cite{Ishkhanyan2019}, and for vanishing spatial scalar curvature, it reduces to Leaver's generalised spheroidal equation and belongs to the confluent-Heun family
\cite{leaver,ronveaux,slavyanov_lay}.

The paper is organised as follows. Section~\ref{sec:geometry} derives the spatial geometry, its Ricci decomposition, and its embedding in the Friedmann-like four-metric. Section~\ref{sec:spectrum} formulates and analyses the scalar spectral problem. Section~\ref{sec:kernels} constructs the scalar-derived curvature kernels, and Sec.~\ref{sec:numerics} provides their finite-shell numerical illustration. Section~\ref{sec:perturbations} relates the construction to dynamical wormhole and LTB settings, uses a scalar field as a normalisation check, and formulates the projected perturbation equations, including the TT sector. Section~\ref{sec:conclusion} summarises the results. Technical radial identities and numerical details are collected in the appendices.

\section{Spatial geometry and trace-free curvature}\label{sec:geometry}

The purpose of this section is to isolate the geometric quantity responsible for the mode mixing analysed below. Locally, a spherically symmetric three-dimensional Riemannian metric can be written in terms of the proper radial distance as
\begin{equation}\label{eq:metric_chi}
\dd s_3^2=\gamma_{ij}\dd x^i\dd x^j  =\dd\chi^2+\sigma^2(\chi)\dd\Omega^2, \quad\mbox{with}\quad \dd\Omega^2=\dd\theta^2+\sin^2\theta\,\dd\phi^2.
\end{equation}
Its scalar curvature is
\begin{equation}\label{eq:scalar_curvature}
 \Rthree=-\frac{2}{\sigma^2}\left[2\sigma\sigma_{,\chi\chi}  -1  +\left(\sigma_{,\chi}\right)^2 \right],
\end{equation}
where we denote derivatives with respect to $\chi$ as $A_{,\chi}\equiv\dd A/\dd\chi$. We require only that the scalar curvature be spatially constant, \(\Rthree=6\eps\). At this stage, no assumption of constant sectional curvature or maximal symmetry is made. Thus, integration of Eq.~\eqref{eq:scalar_curvature} gives the first integral
\begin{equation}\label{eq:sigma_first_integral}
\left(\sigma_{,\chi}\right)^2=1-\eps\sigma^2-\frac{2M}{\sigma}.
\end{equation} 
Here $M$ is an integration constant and, as such, is independent of any coordinate. This remark will be important when comparing this 3D metric with spatial sections of LTB models. Also, \(M\) measures the departure from maximal symmetry. 

For $\dd\sigma/\dd\chi\neq0$, differentiation of Eq.~\eqref{eq:sigma_first_integral} yields a second-order equation involving both parameters
\begin{equation}\label{eq:sigma_second}
 \sigma_{,\chi\chi}=-\eps\sigma+\frac{M}{\sigma^2}.
\end{equation}
Since both sides are continuous, this relation also extends to regular turning points at which \(\dd\sigma/\dd\chi=0\). Equation~\eqref{eq:sigma_second} separates the constant-curvature contribution, which is linear in \(\sigma\), from the inverse-square contribution generated by \(M\). As shown below, only the latter contributes to the trace-free Ricci tensor. 

Introducing the areal radius \(r=\sigma(\chi)\), the metric becomes
\begin{equation}\label{eq:metric_r}
\dd s_3^2=\frac{\dd r^2}{f(r)}+r^2\dd\Omega^2, \quad\mbox{with}\quad f(r)=1-\eps r^2-\frac{2M}{r}.
\end{equation}
An areal-radius chart covers a connected Riemannian region in which \(f(r)>0\). A simple zero \(f(r_h)=0\) may nevertheless correspond to a regular turning point of the areal radius when described in terms of the proper-distance coordinate \(\chi\). For example, when \(\eps=0\) and \(M>0\), the surface \(r=2M\) is a regular minimal sphere, although it forms the boundary of each areal-radius chart. Eq.~\eqref{eq:metric_r} also coincides with the spatial metric induced on a constant-static-time slice of Schwarzschild--(anti-)de Sitter spacetime, with \(\Lambda=3\eps\). When \(\eps=0\), it reduces to the spatial Schwarzschild geometry: its equatorial two-surface has the Flamm-paraboloid embedding, while its smooth two-sided extension contains the Einstein--Rosen bridge \cite{ER35}. 

The sign of the trace-free curvature is fixed without ambiguity by working in the orthonormal proper-distance frame. Equations~\eqref{eq:sigma_first_integral} and \eqref{eq:sigma_second} give
\begin{align}
\Rthree^{\chi}{}_{\chi} &= -\frac{2}{\sigma}\sigma_{,\chi\chi} = 2\eps-\frac{2M}{r^3},\label{eq:ricci_radial}\\
\Rthree^{\theta}{}_{\theta} =\Rthree^{\phi}{}_{\phi} &=\frac{1-\left(\sigma_{,\chi}\right)^2-\sigma\sigma_{,\chi\chi}}{\sigma^2} =2\eps+\frac{M}{r^3}.\label{eq:ricci_angular}
\end{align}
Hence
\begin{equation}\label{eq:ricci_decomposition}
\Ricthree=2\eps\,\delta^i{}_{j}+{}^{(3)}\bar{\mathcal{R}}^i{}_{j},  \quad\mbox{with}\quad {}^{(3)}\bar{\mathcal{R}}^i{}_{j}=\frac{M}{r^3}\diag(-2,1,1),
\end{equation}
where ${}^{(3)}\bar{\mathcal{R}}^i{}_{j}$ represents the traceless component of the 3D Ricci tensor. The ordering of the diagonal components is \((r,\theta,\phi)\). Taking the trace of Eq.~\eqref{eq:ricci_decomposition} recovers \(\Rthree=6\eps\). When \(M=0\), the Ricci tensor is proportional to the metric and the spatial geometry is locally maximally symmetric. When \(M\neq0\), the scalar curvature remains constant, but the trace-free Ricci tensor is nonzero. Thus \(M\) measures the departure from maximal symmetry without changing the scalar-curvature scale.

For the Friedmann-like four-metric
\begin{equation}\label{eq:four_metric}
 \dd s^2=\dd t^2-a^2(t)\gamma_{ij}\dd x^i\dd x^j,
\end{equation}
where \(\gamma_{ij}\) is the time-independent spatial factor of the four-dimensional warped-product metric, the mixed intrinsic curvature scales as
\begin{equation}\label{eq:physical_pi_scaling}
 \Ricthree[h] =\frac{1}{a^2}\left(2\eps\delta^i{}_{j}+{}^{(3)}\bar{\mathcal{R}}^i{}_{j}[\gamma]\right).
\end{equation}
Here $\Ricthree[h]$ is the mixed intrinsic Ricci tensor of $h_{ij}$; the factor $a^{-2}$ in Eq~\eqref{eq:physical_pi_scaling} refers to this mixed-index form. The physical metric induced on a constant-time hypersurface is \(h_{ij}=a^{2}(t)\gamma_{ij}\). As explained in the previous section, the term \emph{Friedmann-like} refers precisely to this separable time dependence and to the Friedmann form of the isotropic field equations; it does not imply spatial homogeneity or maximal symmetry. The FLRW geometry is recovered only when \(M=0\).

It can be shown that the Einstein tensor of metric \eqref{eq:four_metric} with induced metric given by \eqref{eq:metric_r} takes the form
\begin{eqnarray}
G^0{}_0&=&3\left(H^2+\frac{\epsilon}{a^2}\right),\qquad H=\frac{\dot a}{a},\label{eq:friedmann}\\
G^r{}_r&=&\frac{1}{a^2}\left[2a\ddot a+\dot a^2+\epsilon+\frac{2M}{r^3}\right]\quad G^\theta{}_\theta=G^\phi{}_\phi =\frac{1}{a^2}\left[2a\ddot a+\dot a^2+\epsilon-\frac{M}{r^3}\right].
\end{eqnarray}
The $M$-dependence cancels from the isotropic part, $\tfrac13G^i{}_i=a^{-2}(2a\ddot a+\dot a^2+\epsilon)$, leaving the two ordinary Friedmann-like equations with $\epsilon$ playing the role of the curvature constant. All of the $M$-dependence sits in the trace-free part:
\begin{equation}
\ G^i{}_j-\tfrac13\delta^i{}_jG^k{}_k
=\frac{M}{a^2r^3}\,\mathrm{diag}(2,-1,-1)=-\frac{1}{a^2}{}^{(3)}\bar{\mathcal{R}}^i{}_j.\ 
\label{eq:tracefreeG}
\end{equation} 
In Einstein gravity, Eq.~\eqref{eq:tracefreeG} requires a background anisotropic stress whenever \(M\neq0\). The four-metric therefore cannot be supported by a perfect fluid, irrespective of the choice of \(a(t)\),
\(\rho\), or \(p\). The same trace-free spatial curvature is also responsible for the nonvanishing four-dimensional Weyl curvature.

Although the line element can be written in the standard LTB coordinate form, this correspondence is only kinematical. The mass function inferred from the LTB dust equations, for instance, contains a contribution proportional to \(M a(t)\) and is consequently time-dependent when \(M\neq0\). The geometry is therefore not an LTB dust solution, but rather a separable LTB-like geometry supported by anisotropic stress. Its relation to dynamical wormhole and LTB-like geometries is discussed in Sec.~\ref{sec:perturbations}.


\section{The scalar Laplace--Beltrami spectrum}\label{sec:spectrum}
We now construct the scalar harmonics of the spatial metric \(\gamma_{ij}\). The areal radius \(r\) and the proper radial distance \(\chi\) describe the same geometry, but they expose different aspects of the spectral problem. The areal coordinate gives the radial equation in Sturm--Liouville form and is convenient for projection integrals, whereas proper distance makes regularity at a minimal sphere manifest and converts the equation into a one-dimensional Schr\"odinger problem. We therefore keep the two formulations separate before relating their endpoint data.

\subsection{Areal-radius formulation}
\label{subsec:areal_spectrum}

We adopt the spectral convention
\begin{equation}
\label{eq:eigenvalue_problem}
 \lap Q_{(q)}\equiv\frac{1}{\sqrt{\gamma}} \partial_i\left(\sqrt{\gamma}\,\gamma^{ij}\partial_jQ_{(q)}\right)=-q^2 Q_{(q)},
\end{equation}
with \(q^2\geq0\). In the four-dimensional metric \eqref{eq:four_metric}, the corresponding physical squared wave number is \(q^2/a^2(t)\). Unlike a Fourier wave number in a homogeneous space, \(q\) is defined spectrally: its allowed values and normalisation depend on the global radial domain and on the self-adjoint boundary conditions. Spherical symmetry nevertheless preserves the angular labels and permits the separation
\begin{equation}\label{eq:separation}
 Q_{(q,l,m)}(r,\theta,\phi)=R_{(q,l)}(r)Y_l^m(\theta,\phi),
\end{equation}
where \(l=0,1,\ldots\), \(-l\leq m\leq l\), and \(Y_l^m\) are the ordinary spherical harmonics. The nontrivial geometry is therefore entirely contained in the radial equation
\begin{equation}
\label{eq:radial_divergence}
 \sqrt f\, \left(r^2\sqrt f\,R'\right)' +\left[q^2r^2-l(l+1)\right]R=0,
\end{equation}
where a prime denotes \(\dd/\dd r\), and the subscripts on \(R\) are suppressed when no confusion can arise. Expanding the derivative gives
\begin{equation}
\label{eq:radial_expanded}
 r^2fR''+\left(2rf+\frac{r^2}{2}f'\right)R' +\left[q^2r^2-l(l+1)\right]R=0,
\end{equation}
or, explicitly,
\begin{equation}
\label{eq:radial_polynomial}
 \left(r^2-2Mr-\eps r^4\right)R'' +\left(2r-3M-3\eps r^3\right)R' +\left[q^2r^2-l(l+1)\right]R=0.
\end{equation}
For generic nonzero \(M\) and \(\eps\), this is a five-point generalised Heun equation \cite{olver_nist,Ishkhanyan2019}. When needed, we shall use its explicit Frobenius recurrence about a simple zero of \(f\), given in Appendix~\ref{app:frobenius}. However, closed special-function notation does not, by itself, specify the global spectral problem: the operator domain and endpoint conditions are also required. 

Multiplying Eq.~\eqref{eq:radial_divergence} by \(-1/\sqrt f\) gives the Sturm--Liouville form
\begin{equation}
\label{eq:sturm_liouville}
 -\frac{\dd}{\dd r}\left(pR'\right)+v_lR=q^2wR,
\end{equation}
where
\begin{equation}
 p=r^2\sqrt f,
 \qquad
 v_l=\frac{l(l+1)}{\sqrt f},
 \quad \mbox{and} \quad
 w=\frac{r^2}{\sqrt f}.
\end{equation}
The weight \(w\) is fixed by the spatial volume element \(\dd V=(r^2/\sqrt f)\,\dd r\,\dd\Omega\). After angular normalisation, the radial inner product is therefore
\begin{equation}
\label{eq:radial_inner_product}
 \langle R_1,R_2\rangle_l =\int_I \frac{r^2}{\sqrt f}\,R_1^*(r)R_2(r)\,\dd r,
\end{equation}
where \(I=(a,b)\) is a connected interval on which \(f>0\). Defining the radial operator by \(\mathsf L_lR=w^{-1}[-(pR')'+v_lR]\), Green's identity gives
\begin{equation}
\label{eq:boundary_form}
 \langle \mathsf L_lR_1,R_2\rangle_l -\langle R_1,\mathsf L_lR_2\rangle_l =\left.p\left(R_1^*R_2'-R_1^{*\prime}R_2\right)\right|_{a}^{b}.
\end{equation}
A symmetric realisation requires the right-hand side of Eq.~\eqref{eq:boundary_form} to vanish for every pair of functions in its domain; self-adjointness requires a maximal domain with this property. On a finite shell with regular endpoints, this can be enforced, for example, by real Dirichlet, Neumann, or Robin conditions at each endpoint. The boundary form is the radial flux associated with the Laplace--Beltrami operator, so its vanishing excludes an uncontrolled flux through the boundary; a self-adjoint choice then has a real spectrum. A finite-shell realisation in this regard is presented in Sec.~\ref{sec:numerics}, with the numerical implementation described in Appendix~\ref{app:numerics}. At a zero of $f$, \(p\to0\) does not by itself make the boundary form vanish; Sec.~\ref{subsec:endpoints} explains why.

\subsection{Proper-distance and Schr\"odinger formulation}
\label{subsec:schrodinger}

The proper-distance coordinate removes the apparent degeneration of the areal coordinate at a regular turning point. In terms of \(\chi\), the same radial equation is now
\begin{equation}
\label{eq:radial_chi}
 \frac{1}{\sigma^2}\frac{\dd}{\dd\chi} \left(\sigma^2\frac{\dd R}{\dd\chi}\right) -\frac{l(l+1)}{\sigma^2}R=-q^2R.
\end{equation}
Setting
\begin{equation}
\label{eq:u_definition}
 R(\chi)=\frac{u(\chi)}{\sigma(\chi)}
\end{equation}
removes the first derivative. Using Eq.~\eqref{eq:sigma_second}, one obtains
\begin{equation}
\label{eq:schrodinger}
 -\frac{\dd^2u}{\dd\chi^2}+V_l(\chi)u=(q^2+\eps)u,
\end{equation}
with
\begin{equation}
\label{eq:effective_potential}
 V_l(\chi)=\frac{l(l+1)}{\sigma^2(\chi)} +\frac{M}{\sigma^3(\chi)}.
\end{equation}
The radial norm simplifies to
\begin{equation}\label{eq:u_norm}
 \langle R_1,R_2\rangle_l=\int u_1^*(\chi)u_2(\chi)\,\dd\chi.
\end{equation}
Thus the Sturm--Liouville weight in areal radius is converted into the standard \(L^2(\dd\chi)\) norm. The explicit \(\eps\)-dependence appears as the uniform shift from \(q^2\) to the Schr\"odinger eigenvalue \(E=q^2+\eps\), while \(\eps\) also affects the potential implicitly through \(\sigma(\chi)\). The additional term \(M/\sigma^3\) is the spatially varying contribution generated by the trace-free Ricci curvature. For \(M>0\), it is a repulsive geometric contribution that supplements the centrifugal barrier and suppresses radial penetration toward the minimal sphere. In the asymptotically flat case \(\eps=0\), it falls off one power of \(\sigma\) faster than the centrifugal term and therefore modifies scattering phases without changing the asymptotic wave number.

\begin{figure}[t]
 \centering
 \includegraphics[width=1.0\textwidth]{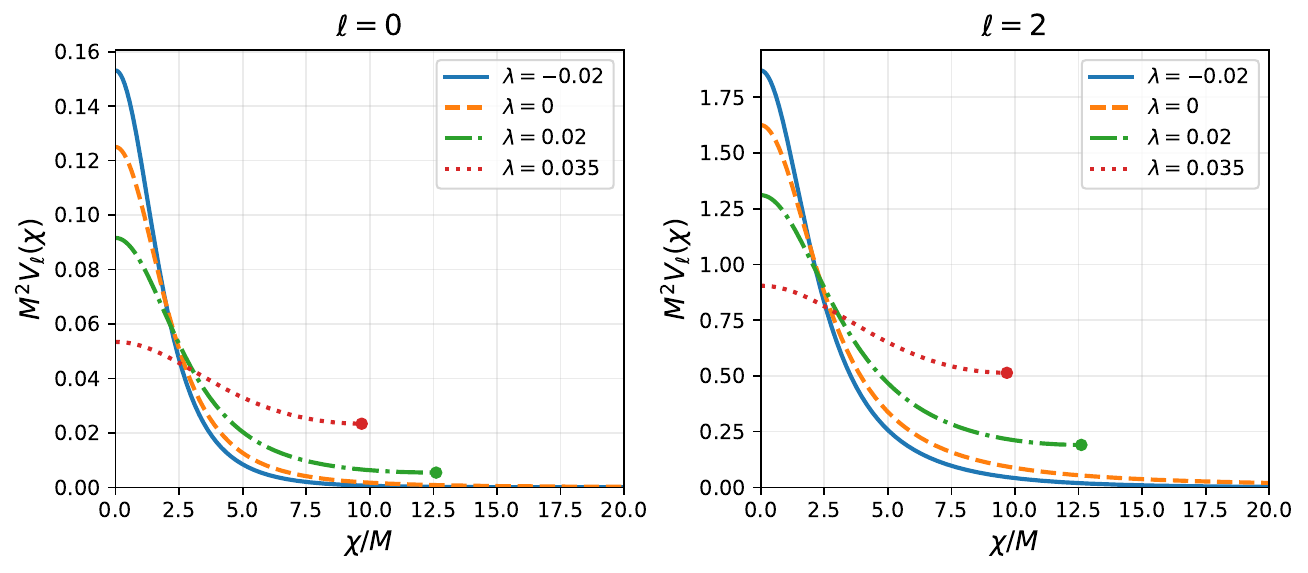}
 \caption{Dimensionless effective potential $M^2V_l$ for $M>0$ and $\lambda=\epsilon M^2=-0.02,0,0.02,0.035$. The panels show $l=0$ (left) and $l=2$ (right). Proper distance is measured from each geometry's minimal sphere. Dots mark the regular maximum sphere reached for $\lambda>0$, where the plotted areal-radius chart ends.}
 \label{fig:radial_potential}
\end{figure}

Equation \eqref{eq:schrodinger} is the standard radial Schr\"odinger equation \cite{AzadLoranMostafazadeh2020}, whose unshifted potential is $l(l+1)/\sigma^2+\sigma_{,\chi\chi}/\sigma$; here it equals $V_l-\epsilon$. Figure~\ref{fig:radial_potential} compares the exact potential for several values of the combination $\lambda\equiv\epsilon M^2$, with proper distance measured from each geometry's minimal sphere. Changing $M$ alone at $\epsilon=0$ leaves the dimensionless solution curves unchanged. For $\epsilon\ne0$, we integrate Eq.~\eqref{eq:sigma_second} with $\sigma(0)=r_{\min}$ and $\sigma_{,\chi}(0)=0$, where $r_{\min}$ is the smallest positive zero of $f$. For $\epsilon=0$, the analytic transformation is
\begin{equation}
\label{eq:proper_distance_exact}
 \chi(\sigma)=\sqrt{\sigma(\sigma-2M)} +2M\operatorname{arcosh}\sqrt{\frac{\sigma}{2M}}.
\end{equation}

\subsection{Endpoint regularity and self-adjointness}
\label{subsec:endpoints}

Let \(r_h\) be a simple zero of \(f\). Although the areal-coordinate equation is singular there, the proper-distance geometry can remain regular. In Eq.~\eqref{eq:radial_expanded}, the coefficient of \(R''\) has a simple zero at \(r_h\), while
\begin{equation}
 \left(2rf+\frac{r^2}{2}f'\right)_{r_h} =\frac12\left(r^2f\right)'_{r_h}.
\end{equation}
The Frobenius indicial equation is therefore
\begin{equation}
\label{eq:indicial}
 s\left(s-\frac12\right)=0,
\end{equation}
so the two local branches have the form
\begin{align}
 R_1(r)&=a_0+a_1(r-r_h)+\cdots,\label{eq:branch_even}\\
 R_2(r)&=|r-r_h|^{1/2}\left[b_0+b_1(r-r_h)+\cdots\right].\label{eq:branch_odd}
\end{align}
Appendix~\ref{app:frobenius} records the recurrence determining all higher coefficients. Both branches are locally square integrable because \(w\sim|r-r_h|^{-1/2}\). Moreover, \(p\sim|r-r_h|^{1/2}\): although \(pR_1'\to0\), the product \(pR_2'\) generally approaches a nonzero constant. This explains why \(p(r_h)=0\) does not automatically make the boundary form \eqref{eq:boundary_form} vanish.

For \(\eps=0\) and \(M>0\), choose \(\chi=0\) at \(r=2M\). Then,
\begin{equation}
\label{eq:throat_expansion}
 \sigma(\chi)=2M+\frac{\chi^2}{8M}+O(\chi^4),
\end{equation}
and
\begin{equation}
\label{eq:potential_throat}
 V_l(0)=\frac{l(l+1)}{4M^2}+\frac{1}{8M^2}.
\end{equation}
Equations~\eqref{eq:branch_even} and \eqref{eq:branch_odd} are simply the even and odd regular Taylor branches in \(\chi\), because \(r-2M=\chi^2/(8M)+O(\chi^4)\). On a smooth two-sided extension, \emph{throat-reflection parity} selects \(\dd u/\dd\chi|_{0}=0\) for the even sector or \(u(0)=0\) for the odd sector. Since $\sigma(\chi)$ is even and $u=\sigma R$, $u$ and $R$ have the same throat-reflection parity. This terminology is distinct from angular polar/axial parity. These are geometric regularity conditions, not artificial walls. If only one exterior is retained, \(\chi=0\) is a limit-circle endpoint and one real Robin condition is instead required to define a self-adjoint realisation. The choice specifies how the one-sided exterior is completed at its inner boundary.

At large radius, with $\eps=0$, we have
\begin{equation}
\label{eq:chi_asymptotic}
\chi(r)=r+M\ln\left(\frac{r}{M}\right)+O(1),
\end{equation}
and \(V_l\to0\). The endpoint at infinity is a limit point, and the generalised eigenfunctions have the scattering form
\begin{equation}
 u(\chi)\sim A_+e^{\ii q\chi}+A_-e^{-\ii q\chi}.
\end{equation}
For \(\eps=0\) and \(M>0\), the potential satisfies \(V_l\geq0\). With either geometrically regular parity condition at the minimal sphere, the corresponding Schr\"odinger operator is nonnegative and therefore has no negative eigenvalues. Since \(V_l(\chi)\to0\) as \(\chi\to\infty\), its essential spectrum is \(q^2\in[0,\infty)\). An arbitrary Robin condition on a one-sided exterior need not preserve nonnegativity and may introduce an isolated negative eigenvalue. Such a bound state would belong to a separate spectral sector; it is absent for the two regular parity extensions considered here. The oscillatory continuum modes are generalised eigenfunctions normalised by Dirac deltas rather than ordinary square-integrable functions. Completeness follows from the spectral theorem only after the self-adjoint domain has been fixed; it does not follow from a local Heun representation alone. For $\epsilon<0$, the $f>0$ chart extends from one regular minimum to a noncompact end. For $\epsilon>0$, an open $f>0$ chart exists only when $0<27\epsilon M^2<1$ and lies between a regular minimum and a regular maximum. Its global spectrum requires a specified completion or boundary conditions at both turning spheres. The double-root case $27\epsilon M^2=1$ is outside the simple-root analysis.

\subsection{Maximally symmetric and exact special cases}
\label{subsec:maximal_limit}

When \(M=0\), Eq.~\eqref{eq:radial_expanded} reduces to the standard radial equation on the maximally symmetric
Robertson--Walker spatial sections. In the flat case, \(\eps=0\),
\begin{equation}
 \frac{\dd}{\dd r}\left(r^2R'\right) +\left[q^2r^2-l(l+1)\right]R=0,
\end{equation}
with solutions
\begin{equation}
\label{eq:flat_bessel}
 R_{ql}^{(0)}(r)=A_{ql}j_l(qr)+B_{ql}y_l(qr),
\end{equation}
where $j_l$ and $y_l$ are spherical Bessel functions of first and second type, respectively. Regularity at the origin selects \(j_l\). For \(\eps=+1\) on the unit three-sphere, compactness makes the spectrum discrete, \(q^2=n(n+2)\), with \(n=0,1,\ldots\) and \(0\leq l\leq n\). The flat space has a continuum beginning at \(q^2=0\), while the \(L^2\) spectrum on the unit open three-space is continuous and begins at \(q^2=1\). These are the standard complete bases discussed in Ref.~\cite{kalnins_miller}.

For the case \(\eps=0\) and \(M>0\), Eq.~\eqref{eq:radial_polynomial} becomes
\begin{equation}
\label{eq:radial_epsilon_zero}
 r(r-2M)R''+(2r-3M)R' +\left[q^2r^2-l(l+1)\right]R=0.
\end{equation}
This equation can be mapped to Leaver's generalised spheroidal equation \cite{leaver}. With the parametrisation
\begin{equation}
 x(x-x_0)y''+(B_1+B_2x)y' +\left[\omega^2x(x-x_0)-2\eta\omega(x-x_0)+B_3\right]y=0,
\end{equation}
the proper identification is
\begin{gather}
\label{eq:leaver_parameters}
 x=r,\qquad x_0=2M,\qquad B_1=-3M,\qquad B_2=2,\\
 \omega=q,\qquad \eta=-Mq,\qquad B_3=-l(l+1)+4M^2q^2.
\end{gather}
Convergent Jaff\'e-type series therefore provide exact local and exterior representations in the confluent-Heun class. The elementary Frobenius series and its recurrence are given in Appendix~\ref{app:frobenius}.

\subsection{Weak-field expansion}
\label{subsec:weak_field}

The dimensionless parameter controlling the local departure from the flat radial operator is \(M/r\). On a shell satisfying \(M/r\ll1\), away from the minimal sphere, Eq.~\eqref{eq:radial_epsilon_zero} may be written
\begin{equation}
\label{eq:weak_operator}
 \mathcal{L}_0[R]=M\mathcal{S}[R],
\end{equation}
where
\begin{align}
 \mathcal{L}_0[R] &\equiv\frac{\dd}{\dd r}\left(r^2R'\right) +\left[q^2r^2-l(l+1)\right]R,\label{eq:L0}\\
 \mathcal{S}[R]&\equiv2rR''+3R'.\label{eq:source}
\end{align}
Writing the solution perturbatively
\begin{equation}
 R=R^{(0)}+MR^{(1)}+O(M^2),
\end{equation}
we have, for the first order
\begin{equation}
\label{eq:first_order_equation}
 \mathcal{L}_0[R^{(0)}]=0, \qquad \mathcal{L}_0[R^{(1)}]=2rR^{(0)\prime\prime}+3R^{(0)\prime}.
\end{equation}
For real \(q\), choose the zeroth-order fundamental pair $J_l(r)\equiv j_l(qr)$ and $N_l(r)\equiv y_l(qr)$. Their radial Wronskian is \(W[J_l,N_l]=1/(qr^2)\). At fixed \(q\), variation of parameters gives the particular solution
\begin{equation}
\label{eq:first_order_integral}
 \begin{aligned}
 R^{(1)}(r)={}&q\Bigg[ -J_l(r)\int_{r_0}^{r}N_l(s)\mathcal{S}[R^{(0)}](s)\,\dd s +N_l(r)\int_{r_0}^{r}J_l(s)\mathcal{S}[R^{(0)}](s)\,\dd s \Bigg]\\
 &+C_JJ_l(r)+C_NN_l(r),
 \end{aligned}
\end{equation}
where the constants, and thus the integration limits, are fixed by the chosen boundary conditions. The correction changes the radial amplitude and phase but introduces no new spectral variable. Using the zeroth-order equation, its source can be reduced to
\begin{equation}
\label{eq:source_reduced}
 \mathcal{S}[R^{(0)}] =-R^{(0)\prime}-2q^2rR^{(0)} +\frac{2l(l+1)}{r}R^{(0)}.
\end{equation}

Equation~\eqref{eq:first_order_integral} is most directly interpreted for continuum scattering at fixed \(q\). On a finite shell, the eigenvalues also shift with \(M\), and standard Sturm--Liouville perturbation theory gives the same result. It is worth mentioning that the eigenvalue shift is not needed for the leading scalar curvature kernel, as we shall see in Sec.~\ref{sec:kernels}: the kernel already contains one explicit power of \(M\), so corrections to the modes and measure first contribute there at \(O(M^2)\), as shown in Appendix~\ref{app:identities}.

At strictly zero spectral parameter, the decaying multipole has the simple asymptotic expansion
\begin{equation}
\label{eq:zero_mode_expansion}
 R_{q=0,l}(r)=\frac{A_l}{r^{l+1}} \left[1+\frac{(2l+1)M}{2r} +O\!\left(\frac{M^2}{r^2}\right)\right].
\end{equation}
For nonzero \(q\), Eq.~\eqref{eq:first_order_integral} supplies the corresponding correction without analytic continuation. If a simultaneous long-wavelength expansion is made, the ordinary \(O(q^2)\) terms and the mixed \(O(Mq^2)\) terms must be tracked consistently.


\section{Scalar-derived harmonics and curvature kernels}\label{sec:kernels}

The contractions generated by the trace-free spatial Ricci tensor are the part of the geometric perturbation operator that is not, in general, diagonalised by the scalar Laplace--Beltrami basis. This section turns that statement into a precise spectral construction. The central object is a kernel whose entry at \((q',q)\) is the projection of a
curvature-distorted input mode with radial label \(q\) onto an output mode with label \(q'\). Thus the kernel is a matrix in radial mode space, or an integral kernel for a continuous spectrum, unlike a Green function, which is defined in position space.

\subsection{Scalar-derived fields and the curvature operator}

Let \(\nabla_i\) be the covariant derivative compatible with \(\gamma_{ij}\), and use \(\gamma_{ij}\) and its inverse to lower and raise spatial indices. For an eigenfunction with \(q>0\), define the scalar-derived covector and trace-free Hessian by
\begin{equation}
\label{eq:derived_harmonics}
 Q_i=\nabla_iQ,  \qquad  \widehat Q^i{}_{j} =\frac{1}{q^2}\nabla^i\nabla_jQ+\frac13\delta^i{}_{j}Q.
\end{equation}
The factor \(q^{-2}\) gives \(\widehat Q^i{}_j\) the same normalisation dimension as \(Q\), but it also means that the zero mode must be treated separately. Thus, if a \(q=0\) sector is present, one should work first with the unnormalised
trace-free Hessian \(\nabla^i\nabla_jQ-\delta^i{}_j\nabla^2Q/3\). 

Commuting covariant derivatives on \(Q_i\) gives
\begin{equation}
\label{eq:divergence_Qhat}
\nabla_j\widehat Q^j{}_{i} =\left(-\frac23+\frac{2\eps}{q^2}\right)Q_i +\frac{1}{q^2}{}^{(3)}\bar{\mathcal{R}}_i{}^jQ_j.
\end{equation}
The first term is the familiar maximally symmetric result. The second term measures its failure when \(M\neq0\).  In an orthonormal radial--tangential frame, the trace-free curvature has eigenvalues \((-2,1,1)M/r^3\). It is therefore spherically symmetric but locally anisotropic: it distinguishes radial from tangential deformations, as a tidal tensor does. Equation~\eqref{eq:divergence_Qhat} shows where this local anisotropy enters the scalar-derived harmonic identities.

To define its spectral matrix elements without distributional normalisations, first place the radial problem on a finite interval with self-adjoint boundary conditions and write
\begin{equation}
\label{eq:scalar_normalization}
\int \dd^3x\sqrt\gamma\, Q_{nlm}^*Q_{n'l'm'} =\calN_{nl}\delta_{nn'}\delta_{ll'}\delta_{mm'}.
\end{equation}
The formulas below are written within a fixed radial channel of the chosen self-adjoint domain. A complete basis on the smooth two-sided $\epsilon=0$ geometry includes both even and odd throat-reflection channels; the reflection-invariant curvature operator preserves each. The continuum formulas follow by replacing the discrete radial index by \(q\), the sum by the appropriate spectral integral, and the Kronecker delta by a Dirac delta. On the nonzero scalar spectrum, it is useful to regard the
contraction as the operator 
\[
\mathsf C ={}^{(3)}\bar{\mathcal{R}}_i{}^j \nabla^i\nabla_j(-\nabla^2)^{-1}.
\]
Because \({}^{(3)}\bar{\mathcal{R}}^i{}_i=0\), its action on an eigenmode is precisely \(\mathsf C Q_{qlm}={}^{(3)}\bar{\mathcal{R}}_i{}^j\widehat Q^i{}_{j,qlm}\). The operator-theoretic criterion for radial spectral mixing is noncommutativity. On a common invariant domain one generally has
\[
[\nabla^ 2, \mathsf C]\neq 0
\]
so Laplace-Beltrami eigenfunctions need not diagonalise $\mathsf C$. Radial inhomogeneity alone is not sufficient: an operator that is a function of $\nabla^2$ remains diagonal in this basis even on the present inhomogeneous geometry.

Expanding the result in the same scalar basis defines
\begin{equation}
\label{eq:a_expansion}
 {}^{(3)}\bar{\mathcal{R}}_i{}^j\widehat Q^i{}_{j,qlm} =\int\dd q'\,a_l(q',q)Q_{q'lm},
\end{equation}
where, for a \(\dd q'\)-normalised continuum basis,
\begin{equation}
\label{eq:a_projection}
 a_l(q',q)=\frac{1}{\calN_{q'l}} \int\dd^3x\sqrt\gamma\, Q_{q'lm}^*{}^{(3)}\bar{\mathcal{R}}_i{}^j
 \widehat Q^i{}_{j,qlm}.
\end{equation}
Thus \(a_l(q',q)\) is simply the \((q',q)\) matrix element of \(\mathsf C\). For example, if a field has radial amplitudes \(F_l(q)\), this part of the operator produces the output amplitude
\[
 (\mathsf C F)_l(q')=\int\dd q\,a_l(q',q)F_l(q).
\]
This equation is the mathematical meaning of the kernel. Spherical symmetry makes the angular projection diagonal in \(l,m\), whereas there is no radial translation symmetry that would force \(a_l(q',q)\) to be proportional to \(\delta(q'-q)\).

\subsection{Radial form of the scalar kernel}

By using the basis decomposition, the radial equation \eqref{eq:radial_expanded}, and the radial and angular Hessian identities given in Eqs.~\eqref{eq:hessian_rr}--\eqref{eq:hessian_angular_trace} of Appendix~\ref{app:identities}, the contraction in the integrand of Eq.~\eqref{eq:a_projection} can be written without second derivatives, as follows
\begin{align}
\label{eq:pi_hessian}
 {}^{(3)}\bar{\mathcal{R}}_i{}^j\nabla^i\nabla_jQ =\frac{M}{r^3}\left[\frac{6f}{r}R' +\left(2q^2-\frac{3l(l+1)}{r^2}\right)R \right]Y_l^m.
\end{align}
It also has a transparent proper-distance form given by
\[
 {}^{(3)}\bar{\mathcal{R}}_i{}^j\nabla^i\nabla_jQ =\frac{M}{\sigma^3}\left[ 6\frac{\sigma_{,\chi}}{\sigma}\,R_{,\chi}
 +\left(2q^2-\frac{3l(l+1)}{\sigma^2}\right)R \right]Y_l^m.
\]
At a regular minimal sphere, one has \(\sigma_{,\chi}=0\). Hence, the derivative contribution vanishes there, and the contraction remains finite. This makes clear that the apparent degeneration of the areal coordinate at \(f=0\)
does not produce a singular curvature coupling.

Since the trace part of \(\widehat Q^i{}_j\) drops out, inserting Eq.~\eqref{eq:pi_hessian} into Eq.~\eqref{eq:a_projection} gives the radial kernel
\begin{align}
\label{eq:a_radial}
 a_l(q',q)=\frac{M}{q^2\calN_{q'l}} \int_I\frac{\dd r}{r\sqrt f}\,R_{q'l}^*(r) \left[ \frac{6f}{r}R_{ql}'(r)
 +\left(2q^2-\frac{3l(l+1)}{r^2}\right)R_{ql}(r) \right].
\end{align}
This equation, including its inverse-Laplacian interpretation, is defined on the $q\neq0$ scalar spectral subspace. Any $q = 0$ sector must instead be treated with the unnormalised trace-free Hessian introduced below Eq.~\eqref{eq:derived_harmonics}. The radial equation cannot, in general, eliminate the remaining first derivative: doing so by integration by parts transfers the derivative to \(R_{q'l}^*\) and produces an endpoint term whose value depends on the chosen boundary conditions. Equation~\eqref{eq:a_radial} is therefore the simplest local form that introduces no additional boundary term.

The operator \(\mathsf C\), and its finite-shell matrix elements, have the dimensions of spatial curvature; the dimensions assigned to a continuum kernel also depend on the chosen spectral measure. In either convention, the kernel carries an explicit factor of \(M\). It vanishes in the maximally symmetric limit and is largest, locally, where \(M/r^3\) is largest, although the sign and size of each matrix element also depend on oscillatory mode overlap. At leading order in \(M\), the modes and measure in Eq.~\eqref{eq:a_radial} may be replaced by their \(M=0\) values, because their first corrections would multiply the explicit factor of \(M\) and contribute only at \(O(M^2)\); this point is detailed in Appendix~\ref{app:identities}. 

\subsection{Vector and tensor projections}

The same geometric distinction between radial and tangential directions is already visible in the vector obtained from \(Q_i\):
\begin{align}
 {}^{(3)}\bar{\mathcal{R}}_r{}^rQ_r&=-\frac{2M}{r^3}R_{ql}'Y_l^m, \label{eq:vector_radial}\\
 {}^{(3)}\bar{\mathcal{R}}_A{}^BQ_B&=\frac{M}{r^3}R_{ql}\mathscr{D}_A Y_l^m, \label{eq:vector_angular}
\end{align}
where \(\mathscr{D}_A\) is the derivative on the unit two-sphere. A pure gradient fixes a relation between its radial and tangential amplitudes. The unequal curvature eigenvalues in Eqs.~\eqref{eq:vector_radial} and \eqref{eq:vector_angular}, together with the radial dependence of \(M/r^3\), do not preserve that relation. A general polar vector at fixed \((l,m)\) therefore requires two independent projections, which may be denoted \(b_l^{\parallel}(q',q)\) and \(b_l^{\perp}(q',q)\). For \(l=0\) the tangential component is absent; for \(l\geq1\) both components are needed.

The same caution applies one rank higher. A product such as \({}^{(3)}\bar{\mathcal{R}}_i{}^k\widehat Q^j{}_k\) generally populates several scalar-parity tensor components and must be projected onto a complete polar tensor basis.  A single unspecified ``tensor coefficient'' cannot describe this action. It is important to mention that this is a statement about basis closure; by itself it does not establish a dynamical conversion between scalar and tensor degrees of freedom.

The resulting selection rules can now be stated precisely. Rotational symmetry and angular parity preserve $l$, $m$,
and the polar/axial sector \cite{GerlachSengupta1979}, whereas the radially varying curvature and derivative contractions can mix radial spectral labels. Thus, the curvature carries no angular momentum and cannot change \(l\), \(m\), or parity, but it can redistribute radial spectral content. The off-diagonal entries quantify this geometric redistribution. They become physical mode couplings only when the perturbation equations contain the corresponding curvature operator; their dynamical effect then also depends on the remaining terms in those equations.

\section{Numerical illustration of radial mode mixing}\label{sec:numerics}

This section first illustrates radial non-diagonality through the common curvature multiplier and then benchmarks one off-diagonal entry of the full scalar kernel \eqref{eq:a_radial}, comparing the leading flat-reference approximation with the exact geometry and eigenfunctions. The multiplier is
\begin{equation}
\label{eq:WM}
\mathcal{W}_M(r)=\frac{M}{r^3},
\end{equation}
which appears both in \({}^{(3)}\bar{\mathcal{R}}^i{}_j\) and in the exact Schr\"odinger potential \eqref{eq:effective_potential}. The numerical question is whether multiplication by this nonconstant function is diagonal in a radial spectral basis. A nonzero off-diagonal matrix element provides a direct illustration of the radial mixing mechanism, although it is not, by itself, the complete kernel \(a_l(q',q)\) of Eq.~\eqref{eq:a_radial}. Accordingly, the multiplier calculation is used only to display the mechanism and its radial selection structure; it is not a quantitatively reliable surrogate for the full derivative-containing kernel.

It is important to state the coordinate choice explicitly. The matrix below is a leading-order calculation about the flat, maximally symmetric reference geometry, \(M=0=\eps\). On that reference geometry, \(f=1\), the areal radius and proper radial distance coincide, \(\chi=r\), and the transformation \(u=rR\) gives
\[
\int_{r_-}^{r_+}r^2R_{nl}^{(0)}R_{n'l}^{(0)}\,\dd r =\int_{r_-}^{r_+}u_{nl}^{(0)}u_{n'l}^{(0)}\,\dd r.
\]
Thus, in this subsection only, writing \(u_{nl}^{(0)}(r)\) does not mix the areal-radius and proper-distance representations: the two coordinates are identical for the reference operator. The exact $M>0$ relation $\chi(r)$ is not used to obtain the flat-reference modes or the multiplier matrix in Fig.~\ref{fig:mode_mixing} and Table~\ref{tab:convergence}. This separation is essential. Since \(\mathcal W_M\) already contains one explicit power of \(M\), replacing the reference modes and measure by their exact \(M>0\) counterparts changes this particular matrix only at \(O(M^2)\), as explained in Appendix~\ref{app:identities}.

We regularise the reference spectrum on the finite shell \(r\in[r_-,r_+]\) and impose Dirichlet conditions. This transparent self-adjoint realisation converts the continuum into a discrete set of modes; the two boundaries are a numerical regulator, not physical walls. Individual matrix elements depend on the shell and boundary conditions, while diagnostics formed from truncated matrices also depend on the retained mode block; the regulator-independent conclusion is the failure of the curvature operator to be diagonal in the Laplace-Beltrami basis. The reference problem is
\begin{equation}
\label{eq:H0_numerical}
H_l^{(0)}=-\frac{\dd^2}{\dd r^2}+\frac{l(l+1)}{r^2}, \qquad u_{nl}^{(0)}(r_-)=u_{nl}^{(0)}(r_+)=0.
\end{equation}
In the flat-reference calculation, \(q_{nl}\) denotes the wave number of \(H_l^{(0)}\). The exact wave numbers used in the coordinate-consistency and full scalar-kernel checks instead belong to the \(M>0\) operator. The reference modes satisfy
\begin{equation}
 H_l^{(0)}u_{nl}^{(0)}=q_{nl}^2u_{nl}^{(0)}, \qquad \int_{r_-}^{r_+}u_{nl}^{(0)}u_{n'l}^{(0)}\dd r=\delta_{nn'}.
\end{equation}
In this basis, the matrix of the isolated multiplier is
\begin{equation}
\label{eq:C_matrix}
C_{nn'}^{(l)} =M\int_{r_-}^{r_+} \frac{u_{nl}^{(0)}(r)u_{n'l}^{(0)}(r)}{r^3}\,\dd r.
\end{equation}
The matrix is real and symmetric. It would be diagonal if \(\mathcal W_M\) were constant, because the modes are orthonormal. Its off-diagonal entries therefore measure precisely the failure of the radially varying curvature profile to commute with the reference spectral decomposition. Notice that the sign of an individual off-diagonal element
depends on the arbitrary overall signs chosen for the two eigenfunctions; only its magnitude is convention independent.

For comparison between mode pairs, we introduce the dimensionless diagnostic
\begin{equation}
\label{eq:Xi_matrix}
 \Xi_{nn'}^{(l)} =\frac{|C_{nn'}^{(l)}|}{q_{nl}q_{n'l}}.
\end{equation}
The denominator removes the curvature dimension by using the characteristic wave numbers of the two modes. The quantity \(\Xi\) is useful for comparing matrix entries, but it is neither an observable nor a substitute for the complete projected perturbation operator.

\subsection{Discretisation and parameter choices}

We choose the representative quadrupolar sector \(l=2\) and take \(r_-/M=10\) and \(r_+/M=80\). Hence \(M/r\leq0.1\) throughout the shell. This gives a reasonably broad radial interval and a weak isolated curvature multiplier. It does not guarantee comparable accuracy for the full derivative-containing kernel, whose cancellations can amplify basis corrections; the exact-basis checks below quantify this distinction.

The numerical procedure consists of four steps.
\begin{enumerate}
 \item Divide the interval into \(N+1\) equal subintervals, leaving  \(N\) interior grid points. Set the boundary values to zero.
 \item Approximate \(-\dd^2/\dd r^2\) by the centred three-point formula. This produces the real symmetric tridiagonal matrix written explicitly in Appendix~\ref{app:numerics}.
 \item Diagonalise that matrix and retain the lowest 20 eigenpairs. The eigensolver returns Euclidean-normalised vectors \(v_{in}\). The rescaled samples \(u_{nl}^{(0)}(r_i)=v_{in}/\sqrt{\Delta r}\) have the required continuum normalisation.
 \item Evaluate Eq.~\eqref{eq:C_matrix} by the corresponding grid quadrature. The factors of \(\Delta r\) cancel, leaving \(C_{nn'}^{(l)}=M\sum_i v_{in}v_{in'}/r_i^3\).
\end{enumerate}
No interpolation between \(r\) and \(\chi\) occurs in these four steps. The production calculation uses \(N=4000\); we use the sequence \(N=500,1000,2000,4000\) to test convergence.

\begin{figure}[t]
 \centering
 \includegraphics[width=0.96\textwidth]{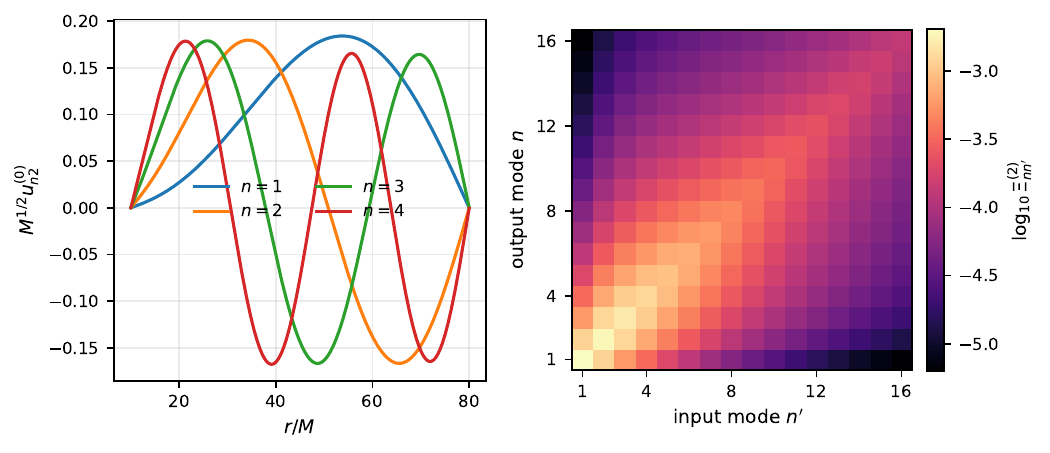}
 \caption{Finite-shell reference calculation for \(l=2\), \(r_-/M=10\), and \(r_+/M=80\).  Left: the first four normalised eigenfunctions of the flat reference operator. On this reference only, \(r=\chi\). Right: the dimensionless diagnostic \(\log_{10}\Xi^{(2)}_{nn'}\). The off-diagonal entries show that the multiplier \(M/r^3\) mixes radial labels while preserving \((l,m)\).}
 \label{fig:mode_mixing}
\end{figure}

Figure~\ref{fig:mode_mixing} shows the first four reference modes and the first \(16\times16\) block of \(\Xi^{(2)}\).  The diagonal band is accompanied by many nonzero off-diagonal entries. Modes with nearby radial orders have the largest overlaps because their oscillation scales are similar; the overlaps decrease as the difference between those scales grows.

\subsection{Convergence and coordinate-consistency checks}

To quantify grid convergence, we monitor the lowest wave number, two matrix elements, and the off-diagonal Frobenius fraction
\begin{equation}
\label{eq:off_fraction}
 F_{\mathrm{off}} =\frac{\|C-\operatorname{diag}C\|_F}{\|C\|_F}
\end{equation}
in the first 12 modes. Table~\ref{tab:convergence} shows that successive errors decrease by approximately a factor of four when \(N\) is doubled. The observed convergence orders obtained from the last three grids are 1.998 for \(q_1\), 1.998 for \(C_{11}\), 1.998 for \(|C_{12}|\), and 2.006 for \(F_{\mathrm{off}}\), in agreement with the expected second-order accuracy of the centred difference formula.

\begin{table}[t]
\caption{Grid convergence of the flat-reference calculation for \(l=2\). The magnitude \(|C_{12}|\), rather than its basis-sign-dependent sign, is reported.}
\label{tab:convergence}
\centering
\begin{tabular}{@{}rcccc@{}}
\toprule
\(N\) & \(q_1M\) & \(M^2C_{11}\) & \(M^2|C_{12}|\) & \(F_{\rm off}\)\\
\midrule
500  & 0.072095328 & \(1.0568069\times10^{-5}\) & \(9.7340113\times10^{-6}\) & 0.85272014\\
1000 & 0.072095407 & \(1.0568122\times10^{-5}\) & \(9.7341240\times10^{-6}\) & 0.85272027\\
2000 & 0.072095427 & \(1.0568136\times10^{-5}\) & \(9.7341522\times10^{-6}\) & 0.85272030\\
4000 & 0.072095432 & \(1.0568139\times10^{-5}\) & \(9.7341592\times10^{-6}\) & 0.85272030\\
\botrule
\end{tabular}
\end{table}

As an independent coordinate-consistency check, the exact \(M>0\), \(\eps=0\) shell problem was solved in two ways: we discretised the Schr\"odinger equation \eqref{eq:schrodinger} on a uniform proper-distance grid, and we separately discretised the generalised Sturm--Liouville problem \eqref{eq:sturm_liouville} on a uniform areal-radius grid. At \(N=4000\), the two calculations give, respectively,
\[
\begin{array}{c|cc}
 & q_1M & M^2C_{11}\\ \hline
\text{proper distance} & 0.0715034442 & 1.0470002875\times10^{-5}\\
\text{areal radius}    & 0.0715034447 & 1.0470002826\times10^{-5}
\end{array}
\]
The relative discrepancies are \(7.0\times10^{-9}\) in \(q_1\) and \(4.7\times10^{-9}\) in \(C_{11}\). This agreement verifies numerically that the two exact coordinate representations describe the same spectral problem. These exact results are a validation check and are not used in Fig.~\ref{fig:mode_mixing} or Table~\ref{tab:convergence}. Their differences from the flat-reference values are $0.83\%$ for $q_1$ and $0.94\%$ for $C_{11}$. These estimates apply to the two reported quantities and do not bound the errors of all modes or of the derivative-containing kernels.

For the first 20 reference modes, the largest off-diagonal entry is \(\max_{n\neq n'}\Xi_{nn'}^{(2)}=1.24837\times10^{-3}\). By contrast, \(F_{\mathrm{off}}=0.8527203\) is comparatively large. There is no contradiction: $F_{\mathrm{off}}$ is a shell-, boundary-, truncation-, and basis-dependent shape statistic for the selected 12-mode block. The overall factor of $M$ cancels from this ratio, so it is not a measure of the absolute physical coupling strength. The small dimensionless entries of \(\Xi\) provide the relevant scale for this particular diagnostic.

Finally, the dependence on the weak-field parameter can be obtained without another numerical run. With \(r=r_-x\), fixed shell ratio \(r_+/r_-\), and normalised modes, dimensional rescaling gives
\begin{equation}
 q_{nl}=r_-^{-1}\widetilde q_{nl},
 \qquad
 C_{nn'}^{(l)}=\frac{M}{r_-^3}\widetilde C_{nn'}^{(l)},
\end{equation}
and therefore
\begin{equation}\label{eq:linear_scaling}
 \Xi_{nn'}^{(l)}=\frac{M}{r_-}
 \frac{|\widetilde C_{nn'}^{(l)}|}
 {\widetilde q_{nl}\widetilde q_{n'l}}.
\end{equation}
Thus, the dimensionless multiplier diagnostic is linear in \(M/r_-\) at leading order. Physically, the broad off-diagonal band arises because the curvature profile is concentrated toward the inner part of the shell and is therefore not adapted to the global radial eigenfunctions. How this geometric mixing enters an actual evolution equation depends on the full kernels of Sec.~\ref{sec:kernels}.

As a check of the full scalar kernel, let \(a^{(l)}_{nn'}\) denote the unit-normalised Dirichlet matrix of Eq.~\eqref{eq:a_radial}, with output index \(n\) and input index \(n'\). For \(\eps=0\), \(l=2\), and \(r\in[10M,80M]\), Table~\ref{tab:scalar_kernel_check} compares the leading flat-reference approximation with the result obtained using the exact geometry and eigenfunctions.

\begin{table}[t]
\caption{Comparison of the off-diagonal scalar-kernel entry
\(M^2|a_{12}^{(2)}|\). The percentage difference is measured
relative to the exact value.}
\label{tab:scalar_kernel_check}
\centering
\begin{tabular}{@{}lcc@{}}
\toprule
Calculation & \(M^2|a_{12}^{(2)}|\) & Difference from exact \\
\midrule
Leading flat-reference approximation
& \(2.71955\times10^{-7}\) & \(-78.4\%\) \\
Exact geometry and eigenfunctions
& \(1.25899\times10^{-6}\) & reference \\
\botrule
\end{tabular}
\end{table}

The exact value is about $4.63$ times the leading result. The \(78.4\%\) discrepancy is a leading-order approximation error: cancellations among derivative contributions amplify changes in the basis and measure, even though the lowest wave number and the isolated multiplier entry \(C_{11}\) differ by less than one per cent between the reference and exact calculations. The multiplier remains useful for demonstrating non-diagonality and selection structure, but for this shell and matrix element, quantitative accuracy requires exact modes or a controlled higher-order approximation. Direct evaluation and an independent integration-by-parts expression agree to a relative Frobenius error below $10^{-10}$ over the first 20 modes in both calculations. The verification code and data are provided in Online Resource 1.

\section{Implications for Friedmann-like perturbations}
\label{sec:perturbations}

The separable four-metric in Eq.~\eqref{eq:four_metric} makes the evolution of a minimal sphere particularly simple. If \(r=r_{\rm th}\) is a minimum of the areal radius on a spatial slice, its physical areal radius is \(R_{\rm th}(t)=a(t)r_{\rm th}\), and hence it expands or contracts at the Hubble rate. This identifies a time-dependent minimal sphere on each constant-\(t\) slice, but it does not by itself establish a traversable wormhole. The causal character of the world tube, the null expansions, the location of trapping horizons, and the global extension must be examined
separately~\cite{bittencourt2018dynamical,bochicchio2010,bronnikov2023,chakraborty2024}.

Equations~\eqref{eq:four_metric} and \eqref{eq:metric_r} can be placed kinematically in the LTB ansatz by taking the LTB areal radius to be \(A(t,r)=a(t)r\). They are not, however, an LTB dust solution: for \(M\neq0\), Eq.~\eqref{eq:tracefreeG} requires a background anisotropic stress. The separability is nevertheless useful because the
harmonics can be constructed once on the time-independent metric \(\gamma_{ij}\); the expansion then enters through \(a(t)\). In a general LTB geometry, \(A(t,r)\) need not be separable, so the spatial operator and its instantaneous eigenbasis are time-dependent, and time derivatives of that basis produce additional mode couplings~\cite{clarkson2009ltb}. The present construction should therefore be regarded as a controlled separable model,
and as a starting point for an adiabatic spectral treatment of more general LTB-like backgrounds.

\subsection{Scalar-field check and the origin of radial mixing}

Because the physical spatial metric is \(h_{ij}=a^2\gamma_{ij}\), its scalar Laplacian is \(a^{-2}\nabla^2\). Expanding a minimally coupled massless scalar field in the complete eigenbasis defined by Eqs.~\eqref{eq:eigenvalue_problem} and \eqref{eq:separation} therefore gives
\begin{equation}
\label{eq:KG_modes}
 \ddot\Phi_{qlm}+3H\dot\Phi_{qlm} +\frac{q^2}{a^2}\Phi_{qlm}=0,
\end{equation}
where \(H\) is defined in Eq.~\eqref{eq:friedmann}. Each radial eigenvalue \(q\) evolves independently: the expansion provides Hubble damping and redshifts the physical wave number to \(q/a\), but it does not mix distinct values of \(q\). Equation~\eqref{eq:KG_modes} is thus both a normalisation check and an important control example. Radial inhomogeneity alone does not produce spectral mixing when the spatial operator depends only on \(\nabla^2\).

Mixing appears when the perturbation equations contain operators that are not diagonal in the Laplace--Beltrami basis, in particular contractions with the trace-free spatial Ricci tensor. After the perturbation variables and matter closure have been specified, any finite-order linear system can be written in first-order form by enlarging the vector of amplitudes as needed. Its curvature-induced part may then be organised schematically as
\begin{equation}
\label{eq:generic_coupled_system}
\dot{\bm{U}}_{qlm}(t) =\mathsf{A}_l(q,t)\bm{U}_{qlm}(t) +\frac{1}{a^2(t)}\int\dd q'\,\mathsf{K}_l(q,q';t)\bm{U}_{q'lm}(t).
\end{equation}
Here \(\bm{U}_{qlm}\) is the column vector of the chosen metric and matter amplitudes, \(\mathsf{A}_l\) contains the terms that are diagonal in \(q\), and \(\mathsf{K}_l\) is the matrix kernel obtained by projecting the curvature-dependent terms. For a discrete radial spectrum, the integral is replaced by a sum. The geometric entries of \(\mathsf{K}_l\) include the scalar kernel in Eq.~\eqref{eq:a_projection}, whose radial form is Eq.~\eqref{eq:a_radial}, as well as the vector and tensor projections described in Sec.~\ref{sec:kernels}. Thus, Eq.~\eqref{eq:generic_coupled_system} defines the projected schematic system, not an additional field equation.

Rotational symmetry makes \(\mathsf{K}_l\) diagonal in \(l,m\), while the radial profile \(M/r^3\) makes it generically off diagonal in \(q\). Here \(\mathsf{K}_l\) denotes only this curvature-induced contribution, which vanishes in the FLRW limit \(M=0\); unrelated matter interactions need not vanish. The factor \(a^{-2}\) follows from the physical curvature scaling in Eq.~\eqref{eq:physical_pi_scaling}, so expansion dilutes its absolute coefficient and contraction enhances it. The diagonal spatial-gradient term has the same scaling, however, so the factor \(a^{-2}\) alone does not make the coupling small relative to that term. Locally the effect is strongest toward the inner region, where \(M/r^3\) is largest. Its dynamical importance must be judged relative to the diagonal frequencies and the Hubble time; the off-diagonal fraction used in Sec.~\ref{sec:numerics} measures the shape of the radial multiplier and not, by itself, the strength of physical mode conversion.

\subsection{Transverse-traceless projection}
\label{subsec:TT_projection}

On a maximally symmetric spatial section, the scalar, vector, and transverse-traceless (TT) tensor sectors are invariant under the linearised spatial operators. This need not remain true when \(M\neq0\), because \(\gamma_{ij}\) is then non-Einstein. In particular, the conditions \(\nabla^i\delta g^{\rm TT}_{ij}=0=\delta g^{{\rm TT}\,i}{}_i\) imposed on initial data do not guarantee that every local curvature term generated during the evolution is TT. One must project the equations and specify the matter-source perturbations; see Ref.~\cite{besse} for the relevant operator framework on non-Einstein manifolds. In Einstein gravity, the background matter cannot be a perfect fluid: as shown by Eqs.~\eqref{eq:physical_pi_scaling} and \eqref{eq:tracefreeG}, it must supply the anisotropic stress required by the trace-free curvature. Its perturbed anisotropic stress depends on the matter model and cannot be inferred from the spatial geometry alone.

Let \(\mathbb{T}^{A}_{q l m\,ij}\) be a complete TT eigenbasis of a chosen self-adjoint reference operator, orthonormal for a discrete spectrum and Dirac-delta normalised with respect to \(\dd q\) for a continuous spectrum. Here \(A\) labels angular polar/axial parity, while \(q\) labels the TT reference spectrum and need not coincide with a scalar Laplace--Beltrami wave number. Let \(\calP_{\rm TT}\) denote the orthogonal TT projection. The projection is understood in the York decomposition \cite{York1973}, with fixed boundary or decay conditions and separate treatment of any null space. The reference operator and its self-adjoint domain are held fixed. We separate the trace-free-curvature contribution to the spatial linearised operator by writing
\begin{equation}
\label{eq:tensor_operator_split}
 \mathcal{O}=\mathcal{O}_{\rm diag} +\delta\mathcal{O}_{{}^{(3)}\bar{\mathcal{R}}}.
\end{equation}
Here \(\delta\mathcal{O}_{{}^{(3)}\bar{\mathcal{R}}}\) is an operator whose local coefficients are linear in
\({}^{(3)}\bar{\mathcal{R}}^i{}_j\); by Eq.~\eqref{eq:ricci_decomposition}, those coefficients scale as \(M/r^3\).
The corresponding geometric TT kernel is the matrix element
\begin{equation}
\label{eq:tensor_kernel}
 \mathcal{M}^{AA'}_l(q,q') =\left\langle \mathbb{T}^{A}_{q l m}, \calP_{\rm TT}\delta\mathcal{O}_{{}^{(3)}\bar{\mathcal{R}}} \mathbb{T}^{A'}_{q'l m} \right\rangle_{\rm TT},
\end{equation}
where the inner product is that induced by \(\gamma_{ij}\) and by the chosen self-adjoint radial domain. Equation~\eqref{eq:tensor_kernel} is the tensor analogue of the scalar projection in Eq.~\eqref{eq:a_projection}: it is the matrix representation of a geometric operator, rather than a new material interaction.

Without a specified perturbation system for the supporting anisotropic matter, neither the kinetic structure nor a universal second-order gravitational-wave equation follows from the geometry alone. Consistently with the schematic system in Eq.~\eqref{eq:generic_coupled_system}, the TT-projected amplitudes can instead be represented in the general first-order form
\begin{equation}
\label{eq:tensor_coupled}
\dot{\bm{U}}^A_{qlm}(t) =\mathsf{A}^A_l(q,t)\bm{U}^A_{qlm}(t) +\frac{1}{a^2(t)}\sum_{A'}\int\dd q'\,\mathsf{K}^{AA'}_l(q,q';t)\bm{U}^{A'}_{q'lm}(t) + \bm{S}^A_{qlm}(t).
\end{equation}
Here $\bm{U}^A_{qlm}(t)$ contains the selected TT and supporting matter amplitudes, enlarged as necessary to obtain first-order form; $\mathsf{A}^A_l$ collects terms diagonal in $q$, and $\bm{S}^A_{qlm}$ denotes retained projected sources or couplings to variables outside this state vector. The geometric TT contribution to the appropriate block of $\mathsf{K}^{AA'}_l$ is supplied by the matrix element $\mathcal{M}^{AA'}_l$ in Eq.~\eqref{eq:tensor_kernel}. Any additional matter kernel is model-dependent and is not fixed by that equation.

Spherical symmetry preserves \(l,m\), and parity-invariant dynamics makes the opposite-angular-parity blocks vanish. The inhomogeneous tidal profile \(M/r^3\), by contrast, can transfer amplitude between different radial eigenfunctions. This is the same geometric mechanism illustrated by the multiplier matrix in Sec.~\ref{sec:numerics}, although that numerical matrix is neither the complete TT kernel nor a quantitative estimate of it. Equation~\eqref{eq:tensor_coupled} is therefore a schematic projected organisation of a model-dependent first-order system, not a derived universal gravitational-wave evolution equation.


\section{Discussion and conclusions}\label{sec:conclusion}

The main lesson of this work is that constant spatial scalar curvature is not enough to retain the usual FLRW mode decomposition. The geometries considered here possess a single scale factor and Friedmann-type isotropic evolution, but
their spatial sections are not maximally symmetric. The parameter \(\eps\) sets the isotropic curvature scale, whereas \(M\) generates trace-free Ricci curvature without changing the spatial Ricci scalar. In the four-dimensional
spacetime, this trace-free part appears as a tidal gravitational field, produces nonzero Weyl curvature, and, within Einstein gravity, requires a supporting anisotropic stress. It is this locally anisotropic curvature that changes the perturbative mode structure.

We have placed the corresponding scalar harmonic problem on a consistent spectral footing. The formulation in areal radius identifies the Sturm--Liouville measure and boundary form needed for self-adjointness, while the formulation in proper distance makes regularity at a minimal sphere transparent and provides a standard one-dimensional Schr\"odinger description. Together, the two representations distinguish coordinate singularities from genuine geometric endpoints and show how the admissible spectrum depends on the global domain and its boundary conditions. The standard Robertson--Walker harmonics are recovered smoothly when the trace-free curvature is removed, whereas the nonmaximally symmetric case retains both an exact special-function description and a controlled weak-field treatment.

The central physical result is a precise form of radial spectral mixing. Spherical symmetry continues to protect the angular momentum labels and, for parity-invariant dynamics, angular polar/axial parity. The radially varying tidal curvature does not share the eigenbasis of the scalar Laplacian, however, and therefore couples modes with different radial spectral labels. The kernels constructed in this paper give the matrix elements of that geometric action. This also clarifies an important limitation: radial inhomogeneity does not make every field equation mode-coupled. A minimally coupled scalar field remains diagonal in the exact Laplace--Beltrami basis. Mixing arises only when the perturbation operator contains curvature contractions, derivative commutators, or other terms that are not functions of the scalar Laplacian alone.

The finite-shell calculation provides a reproducible illustration of this mechanism. It shows a convergent off-diagonal matrix, verifies that the areal-radius and proper-distance descriptions represent the same exact spectral problem, and recovers the expected linear weak-field scaling. Its
physical interpretation is simple: a tidal profile concentrated toward the inner region acts as a radial scattering structure for eigenfunctions that extend across the whole shell. Figure~\ref{fig:mode_mixing} isolates the common curvature multiplier, whereas the additional scalar-kernel check retains the derivative terms of Eq.~\eqref{eq:a_radial}. The size of either off-diagonal matrix does not by itself determine an observable or the total dynamical coupling in a closed perturbation system.

The projected formulation separates the geometric result from assumptions about the matter source. On a non-Einstein spatial background, the usual scalar, vector, and transverse-traceless sectors cannot be assumed to remain independent under every local curvature operator. The relevant tensor
quantity is therefore the transverse-traceless projected kernel. A complete evolution problem must additionally specify the anisotropic matter that supports the background and the perturbation of its stress. This distinction prevents a purely geometric mixing kernel from being mistaken for a universal scalar-to-tensor conversion law, while still identifying the part of the coupling that is fixed by the background geometry.

The same geometry also provides a controlled link with dynamical wormholes and LTB-like spacetimes. In the separable model, a minimal sphere is comoving: its physical areal radius follows the cosmological scale factor, while its area scales with the square of that factor. Likewise, the metric fits the LTB ansatz kinematically but cannot be an LTB dust solution. In a general nonseparable LTB background, the harmonic basis is itself time-dependent and produces further couplings beyond those isolated here.

The novelty of the present analysis is therefore not merely the observation that inhomogeneous curvature can couple perturbations. Earlier work identified the relevant curvature term at the level of the scalar perturbation equations~\cite{santos2015}; here its radial spectral content has been made explicit. The paper combines a self-adjoint harmonic construction, regular endpoint analysis, explicit projection kernels, symmetry selection rules, and a convergent numerical demonstration within one framework. This provides the ingredients needed to move from a formal curvature coupling to model-specific predictions. Natural next steps include choosing a concrete anisotropic matter model, evolving the resulting coupled amplitudes, determining whether the mixing produces measurable changes in stability or propagation, and extending the construction to instantaneous or adiabatic bases in genuinely time-dependent LTB geometries.

\begin{appendices}

\section{Radial identities for the scalar kernel}\label{app:identities}

For \(Q=R(r)Y_l^m(\theta,\phi)\), the mixed Hessian components are
\begin{align}
 \nabla^r\nabla_rQ &=\left(fR''+\frac{f'}{2}R'\right)Y_l^m,\label{eq:hessian_rr}\\
 \nabla^A\nabla_BQ &=\frac{R}{r^2}\mathscr{D}^A\mathscr{D}_BY_l^m +\frac{fR'}{r}\delta^A{}_B Y_l^m. \label{eq:hessian_AB}
\end{align}
Taking the angular trace and using \(\mathscr{D}^2Y_l^m=-l(l+1)Y_l^m\), yields
\begin{equation}
\label{eq:hessian_angular_trace}
 \nabla^A\nabla_AQ =\left[-\frac{l(l+1)}{r^2}R+\frac{2f}{r}R'\right]Y_l^m.
\end{equation}
Contracting Eqs.~\eqref{eq:hessian_rr} and \eqref{eq:hessian_angular_trace} with \({}^{(3)}\bar{\mathcal{R}}^i{}_j=(M/r^3)\diag(-2,1,1)\), and using the radial equation \eqref{eq:radial_expanded} to eliminate the second derivative terms, gives exactly Eq.~\eqref{eq:pi_hessian}.

At fixed $\epsilon$, on a fixed radial domain with fixed boundary conditions, the explicit prefactor in Eq.~\eqref{eq:a_radial} supplies one power
of $M$. Its leading $O(M)$ kernel therefore uses the $M=0$ Robertson--Walker modes at the same $\epsilon$ and the weight $w_0=r^2/\sqrt{1-\epsilon r^2}$, reducing to spherical Bessel functions only for $\epsilon=0$. Corrections to the modes, eigenvalues, and measure enter this kernel at $O(M^2)$, although they enter the scalar spectral operator at first order.

\section{Numerical implementation}\label{app:numerics}

For the flat-reference multiplier calculation in
Sec.~\ref{sec:numerics}, the interval is divided into \(N+1\) equal subintervals of width
\begin{equation}
 \Delta r=\frac{r_+-r_-}{N+1}, \qquad r_i=r_-+i\Delta r, \qquad i=1,\ldots,N.
\end{equation}
The second-order central-difference representation of Eq.~\eqref{eq:H0_numerical} is tridiagonal,
\begin{align}
 (H_l^{(0)})_{ii} &=\frac{2}{\Delta r^2}+\frac{l(l+1)}{r_i^2},\\
 (H_l^{(0)})_{i,i\pm1} &=-\frac{1}{\Delta r^2}.
\end{align}
If \(v_{in}\) are the Euclidean-normalised eigenvectors of this matrix, the continuum-normalised samples are \(u_{nl}^{(0)}(r_i)=v_{in}/\sqrt{\Delta r}\). The quadrature for Eq.~\eqref{eq:C_matrix} then reduces to 
\begin{equation}
 C_{nn'}^{(l)}  =M\sum_{i=1}^{N}\frac{v_{in}v_{in'}}{r_i^3}.
\end{equation}
The convergence sequence in Table~\ref{tab:convergence} is consistent with the expected second-order discretisation error until floating-point and eigensolver tolerances dominate.

For the exact coordinate-consistency check, the
proper-distance calculation uses the same three-point second-derivative stencil with the potential \eqref{eq:effective_potential}. The areal-radius calculation discretises the flux \(pR'\) at cell midpoints and transforms the resulting generalised eigenproblem to a symmetric standard problem using the diagonal matrix of weights \(w(r_i)\). Both calculations impose Dirichlet conditions and retain the lowest 20 modes on grids with \(N=4000\).

The full scalar-kernel comparison in Table~\ref{tab:scalar_kernel_check} uses analytic spherical-Bessel reference modes and independently computed exact modes. In the numerical units \(M=1\), the reference wave numbers are the positive roots of
\[
j_l(qr_+)y_l(qr_-)-y_l(qr_+)j_l(qr_-)=0,
\]
found by Brent's method with absolute tolerance \(10^{-14}\). The reference modes are normalised by \(\int_{r_-}^{r_+}r^2|R_{nl}^{(0)}|^2\,\dd r=1\). Their derivatives are evaluated analytically, and the normalisation and kernel integrals use 1500-point Gauss--Legendre quadrature in \(r\).

For the exact modes, Eqs.~\eqref{eq:sigma_second} and \eqref{eq:schrodinger} are integrated jointly at \(\eps=0\) using DOP853, with relative tolerance \(10^{-12}\) and absolute tolerance \(10^{-13}\). Writing \(\chi_\pm=\chi(r_\pm)\), the shooting solution starts with \(u(\chi_-)=0\) and \(u_{,\chi}(\chi_-)=1\). Finite-difference eigenvalues provide brackets, and Brent's method adjusts \(q\) until \(u(\chi_+)=0\), with absolute root tolerance \(2\times10^{-14}\). The modes are then normalised by \(\int_{\chi_-}^{\chi_+}|u_{nl}|^2\,\dd\chi=1\). Normalisation and kernel integrals use 700-point Gauss--Legendre quadrature in \(\chi\), with \(R=u/r\).

For unit-normalised Dirichlet modes, integration by parts gives the independent expression
\begin{equation}
\label{eq:a_radial_ibp}
a^{(l)}_{nn'}= \frac{M}{q_{n'l}^{2}}\int_{r_-}^{r_+} \left[\frac{2\sqrt f}{r}  \bigl(R_{nl}'\bigr)^*R_{n'l}' -\frac{l(l+1)}{r^3\sqrt f} R_{nl}^*R_{n'l} \right]\dd r.
\end{equation}
The boundary term vanishes under the Dirichlet conditions. The leading-order expression uses \(f=1\) and the reference modes and wave numbers, while retaining the explicit factor of \(M\). The exact expression uses the exact geometry, modes, and wave numbers. For each calculation, the direct and integrated matrices satisfy
\[
\frac{\|a_{\rm direct}-a_{\rm IBP}\|_F}{\|a_{\rm IBP}\|_F}<10^{-10}
\]
on the first 20 modes.

\section{Frobenius recurrence}\label{app:frobenius}

For completeness, write Eq.~\eqref{eq:radial_polynomial} near a simple root \(r_h\) of \(f\) in terms of \(x=r-r_h\):
\begin{equation}
 A(x)R''+B(x)R'+C(x)R=0,
\end{equation}
with
\begin{align}
 A(x)&=\sum_{j=1}^{4}a_jx^j,&  B(x)&=\sum_{j=0}^{3}b_jx^j,& C(x)&=\sum_{j=0}^{2}c_jx^j.
\end{align}
The coefficients are
\begin{align}
 a_1&=2r_h-2M-4\eps r_h^3,& a_2&=1-6\eps r_h^2,& a_3&=-4\eps r_h,& a_4&=-\eps,\\ 
 b_0&=2r_h-3M-3\eps r_h^3=\frac{a_1}{2},& b_1&=2-9\eps r_h^2,& b_2&=-9\eps r_h,& b_3&=-3\eps,\\
 c_0&=q^2r_h^2-l(l+1),& c_1&=2q^2r_h,& c_2&=q^2.
\end{align}
For
\begin{equation}
 R=|x|^s\sum_{n=0}^{\infty}d_nx^n,\qquad d_0\neq0,
\end{equation}
the leading equation is
\begin{equation}
 s\left[a_1(s-1)+b_0\right]=0,
\end{equation}
which gives \(s=0\) and \(s=1/2\). For \(m\geq1\), the coefficients obey
\begin{equation}
 d_m=-\frac{\mathcal{B}_m}{(m+s)[a_1(m+s-1)+b_0]},
\end{equation}
with the numerator given by
\begin{align}
 \mathcal{B}_m={}& \sum_{j=2}^{\min(4,m+1)} a_j(m+s+1-j)(m+s-j)d_{m+1-j}\nonumber\\
 &+\sum_{j=1}^{\min(3,m)}b_j(m+s-j)d_{m-j}+\sum_{j=0}^{\min(2,m-1)}c_jd_{m-1-j}.
\end{align}
Note that coefficients with negative indices are absent. This recurrence constructs both local Frobenius branches about a simple zero of $f$.

\end{appendices}

\backmatter

\section*{Acknowledgements}

E.B. is partially supported by the Conselho Nacional de Desenvolvimento Cient\'ifico e Tecnol\'ogico (CNPq), grant No.~306305/2026-7, and by the Funda\c{c}\~ao de Amparo \`a Pesquisa do Estado de Minas Gerais (FAPEMIG), grant No.~APQ-05207-23. PKSD is supported by a grant from First Rand Bank (SA). E.B. is indebted to Bartolomeu Figueiredo (in memoriam) for valuable comments on an earlier version of this work. This paper is dedicated to his memory.

\subsection*{Author contributions}

All authors contributed to the conceptual development, mathematical interpretation, validation, and review of the manuscript.

\subsection*{Data availability}

All numerical data reported in this study are contained in the article and can be regenerated using the code provided as Online Resource 1. No external data sets were used.

\bibliography{ref}

\end{document}